\documentclass{aastex63}

\usepackage[version=3]{mhchem}
\usepackage{booktabs}
\usepackage{color,soul}
\usepackage{savesym}
\savesymbol{tablenum}
\usepackage{siunitx}
\usepackage{tikz}
\usepackage{enumerate}
\usepackage{float}

\restoresymbol{SIX}{tablenum}
\DeclareSIUnit{\molecule}{molecule}
\definecolor{mypink}{rgb}{0.858, 0.188, 0.478}
\definecolor{mypink3}{cmyk}{0, 0.8808, 0.9429, 0.3412}
\definecolor{mygreen}{cmyk}{0.59, 0.00, 0.99, 0.10}
\definecolor{lightblue}{rgb}{0.659,0.8706,1}
\definecolor{blu}{rgb}{0.2039,0.388,1}
\sethlcolor{lightblue} \setstcolor{red}

\renewcommand{\thefootnote}{\fnsymbol{footnote}}
\renewcommand\arraystretch{1.2}
\submitjournal{ApJL}

\shorttitle{Formation route of propargylimine}
\shortauthors{Lupi et al.}
\graphicspath{{./}{figures/}}

\begin{document}

\title{Methanimine as a key precursor of imines in the interstellar medium: the case of propargylimine}

\correspondingauthor{Vincenzo Barone}
\email{vincenzo.barone@sns.it}
\correspondingauthor{Cristina Puzzarini}
\email{cristina.puzzarini@unibo.it}

\author[0000-0001-6522-9947]{Jacopo Lupi}
\affiliation{Scuola Normale Superiore,
Piazza dei Cavalieri 7,
Pisa, 56126, Italy}

\author[0000-0002-2395-8532]{Cristina Puzzarini}
\affiliation{Department of Chemistry ``Giacomo Ciamician'', University of Bologna,
Via F. Selmi 2,
Bologna, 40126, Italy}

\author[0000-0001-6420-4107]{Vincenzo Barone}
\affiliation{Scuola Normale Superiore,
Piazza dei Cavalieri 7,
Pisa, 56126, Italy}



\begin{abstract}
A gas-phase formation route is proposed for the recently detected propargylimine molecule. In analogy to other imines, such as cyanomethanimine, the addition of a reactive radical (\ce{C2H} in the present case) to methanimine (\ce{CH2NH}) leads to reaction channels open also in the harsh conditions of the interstellar medium. Three possible isomers can be formed in the \ce{CH2NH + C2H} reaction: Z- and E-propargylimine (Z-,E-PGIM) as well as N-ethynyl-methanimine (N-EMIM). For both PGIM species, the computed global rate coefficient is nearly constant in the 20-300 K temperature range, and of the order of 2-3 $\times$ \num{d-10} cm$^3$ molecule$^{-1}$ s$^{-1}$, while that for N-EMIM is about two orders of magnitude smaller. Assuming equal destruction rates for the two isomers, these results imply an abundance ratio for PGIM of [Z]/[E] $\sim$ 1.5, which is only slightly underestimated with respect to the observational datum.
\end{abstract}



\section{Introduction} 
\label{sec:intro}
Currently, the number of molecules detected in the interstellar medium (ISM) thanks to their rotational signatures far exceeds 200 \citep{McGuire_2018}. Among them more than 70 species belong to the class 
of the so-called interstellar complex organic molecules (iCOMs), namely molecules containing at least one carbon atom and a total of more than 6 atoms \citep{COMs}.
Nitrogen-bearing iCOMs are particularly interesting because of their prebiotic character; indeed, they represent key intermediates toward the main building blocks of biomolecules, like aminoacids and nucleobases. Within this class of iCOMs, six members of 
the imine family have been detected so far in the ISM, namely methanimine (\ce{CH2NH}, \cite{godfrey73,Dickens_1997}), ethanimine (\ce{CH3CHNH}, \cite{loomis2013detection}), ketenimine (\ce{CH2CNH}, \cite{Lovas_2006}), 
3-imino-1,2-propadienylidene (\ce{CCCNH}, \cite{kawaguchi1992detection}), C-cyanomethanimine (\ce{NCCHNH}, \cite{zaleski2013detection}; \cite{rivilla2019}), and --very recently-- Z-propargylimine  (2-propyn-1-imine, \ce{HC\bond{=}C\bond{-}CH\bond{=}NH}, \cite{Bizzocchi2020}).

The main hypotheses on their formation mechanisms in astrophysical environments 
involve either tautomerization of simple nitriles \citep{Lovas_2006} or their partial hydrogenation on dust-grain surface \citep{Theule2011,Krim2019}. However, for C-cyanomethanimine, a gas-phase formation route has been recently proposed, which involves addition of the 
cyano radical (CN) to methanimine \citep{Vazart2015}. It is thus 
quite natural to hypothesize that methanimine can play a role in the formation of other imines upon addition/elimination of reactive radicals already detected in the ISM, like \ce{CH3}, \ce{C2H} or \ce{OH}. Indeed, the reaction of the hydroxyl radical with methanimine is proven to effectively lead to the formation of formamide in the gas phase \citep{Vazart2016,formamide-solis}. 

The focus of the present letter is the possible formation pathway of propargylimine (PGIM), whose Z-isomer has been very recently identified in the quiescent G+0.693-0.027 molecular cloud with an estimated column density  of \SI{0.24\pm0.02d14}{\per\square\centi\meter} \citep{Bizzocchi2020}. 
In the same study, an upper limit of \num{1.8d-10} was retrieved for the fractional
abundance (w.r.t. \ce{H2}) of the higher-energy E isomer (which means a column density $<$0.13), which was instead not observed. After the spectroscopic characterization of this imine and its astronomical detection, \cite{Bizzocchi2020} put forward some speculations about 
feasible formation routes based on the relative abundances of a number of possible precursors in the G+0.693-0.027 molecular cloud. However, in spite of the detection of \ce{CH2NH} \citep{Zeng2018}, this has not been taken into consideration, notwithstanding the authors reported, among the others, a large fractional abundance for the ethynyl radical (\ce{C2H}, $^2\Sigma^+$), i.e. \num{3.91d-8}. 

Based on these premises, we decided to perform a state-of-the-art quantum-chemical (QC) characterization of the stationary points on the doublet reactive \ce{C2H + CH2NH} potential energy surface (PES) followed by kinetic computations in the framework of a master equation model rooted in generalized transition state estimates of the elementary reaction rates. 
From a theoretical point of view, the reactions between the ethynyl radical and several substrates have been recently investigated by state-of-the-art QC approaches \citep{Bowman20}, but addition/elimination reactions with unsaturated substrates have not yet been explored.

\section{Computational methodology} 
\label{sec:compdet}
The starting point for the study of the formation pathway of PGIM is the identification of the potential reactants and the analysis of the corresponding reactive PES, which implies the accurate characterization of all stationary points from both a structural and energetic point of view. This first step then requires to be completed by kinetic calculations. 
In the derivation of a feasible reaction mechanism, one has to take into account the extreme conditions of the ISM: low temperatures (10-100 K) and low number density (10-10$^7$ cm$^{-3}$). By translating density in terms of pressure, a number density of 10$^4$ cm$^{-3}$ corresponds to a pressure of 3.8$\times$10$^{-10}$ Pa ($\sim$3.8$\times$10$^{-15}$ atm). 

\subsection{Reactive potential energy surface}
\label{pes:det}
We have followed the general computational strategy validated in several recent studies \citep{earthspace2020,Molecules,lupi:h2s,staa1652,Tonolo2020}, which involves the following steps:
\begin{enumerate}[i)]
\item The stationary points have been located and characterized using the double-hybrid B2PLYP functional \citep{doi:grimme2006}, combined with D3(BJ) corrections (to incorporate dispersion effects; \cite{D3,D3BJ}) and in conjunction with the jun-cc-pVTZ ``seasonal'' basis set \citep{papajak2009}.
\item Single-point energy calculations, at the B2PLYP-D3(BJ)/jun-cc-pVTZ geometries, have been performed by means of the so-called ``cheap'' composite scheme (ChS; \cite{cheap1,cheap2}), which starts from the coupled-cluster theory including single and double excitations augmented by a perturbative estimate of triples (CCSD(T); \cite{Pople89}) in conjunction with a triple-zeta basis set (cc-pVTZ; \cite{Dunning-JCP1989_cc-pVxZ}) and within the frozen-core (fc) approximation. To improve this level of theory, the ChS model considers the extrapolation to the complete basis set (CBS) limit and the effect of core-valence (CV) correlation using M{\o}ller-Plesset theory to second order (MP2; \cite{mp2}). Concerning the former contribution, the fc-MP2 energy is extrapolated to the CBS limit using the $n^{-3}$ expression \citep{Helgaker1997} in conjunction with the cc-pVTZ and cc-pVQZ basis sets. The CV correlation correction is, instead, the difference between the MP2 energy evaluated correlating all electrons and that computed within the fc approximation, both in conjunction with the cc-pCVTZ basis set \citep{cvbasis}. 
\item ChS energies have been combined with anharmonic zero-point energy (ZPE) corrections evaluated at the B2PLYP-D3(BJ)/jun-cc-pVTZ level within hybrid degeneracy-corrected second-order vibrational perturbation theory (HDCPT2; \cite{Bloino2012,chemRev2019}). 
\end{enumerate}

All calculations have been performed with the Gaussian software \citep{g16}.

\subsection{Kinetic models}
\label{Kin:Mod}
Global rate constants have been calculated by using a master equation (ME) approach based on \emph{ab initio}
transition state theory (AITSTME), thereby employing the MESS software as master equation solver
\citep{georgievskii2013reformulation}. For elementary reactions involving a transition state, rate constants
have been computed using transition state theory (TST), while for barrier-less elementary reactions, they have
been evaluated by means of phase space theory (PST; \cite{pechukas1965detailed,chesnavich1986multiple}). The basic assumption of PST is that the interaction between two reacting fragments is isotropic (following a $\frac{C_6}{R^6}$ power law) and does not affect the internal fragment motions \citep{doi:10.1021/cr050205w}. This approximation is generally valid for low-temperature phenomena, as those occurring in the ISM. To be more precise, in order to obtain the $C_6$ parameter for the PST calculation, we performed a scan of the \ce{HCC\bond{-}CH2NH} and \ce{HCC\bond{-}NHCH2} distances for the C- and N-end attack, respectively. Then, the corresponding minimum energy paths have been fitted to a $f(x)$=$f_0 - \frac{C_6}{x^6}$ function, thus obtaining a $C_6$ value of \SI{131.96}{\bohr\tothe{6}\hartree} for the former attack and of \SI{180.59}{\bohr\tothe{6}\hartree} for the latter.
In all cases, tunnelling has been accounted for using the Eckart model \citep{eckart1930penetration}.

The rate constants of the overall reactions leading to the \ce{C3H3N} imine isomers (namely, the E-,Z-PGIM species and N-ethynyl-methanimine, N-EMIM) have been evaluated in the 20-500 K temperature range. To model their temperature dependence, the rate constants at different temperatures have been fitted to a three-parameter modified Arrhenius equation, namely the Arrhenius-Kooij expression \citep{kooij1893zersetzung,laidler1996glossary}:
\begin{equation}
\label{eq:kooij}
    k(T)=A\left(\frac{T}{300}\right)^n\exp\left(-\frac{E}{RT}\right)
\end{equation}
where $A$, $n$, and $E$ are the fitting parameters, $R$ being the universal gas constant.

\begin{figure}[!htbp]
\begin{center}
 \includegraphics[width=1.0\textwidth]{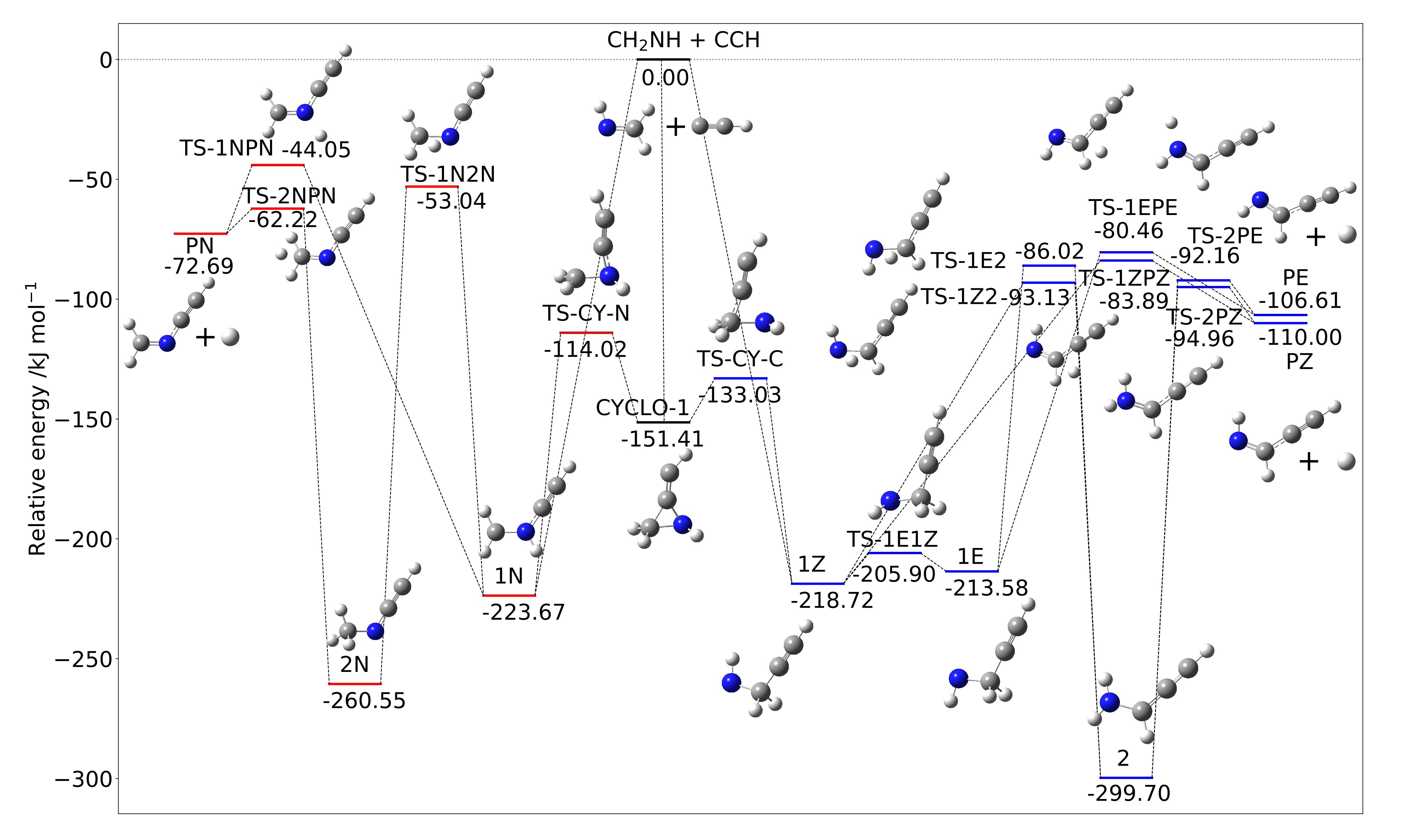}
 \end{center}
 \caption{Formation route of N-EMIM and the PGIM isomers: ChS energies augmented by anharmonic B2PLYP-D3(BJ) ZPE corrections.}
 \label{fig:pespgimchs}
\end{figure}

\section{Results and discussion}

\subsection{Reactivity and energetics}

A recent re-investigation of the reaction channel starting from attack of the cyano radical to the C-end of methanimine \citep{Puzzarini2020} has shown that, for all stationary points, the ChS model has a maximum absolute deviation of 3  \SI{}{\kilo\joule\per\mol} and an average absolute deviation of 1.1  \SI{}{\kilo\joule\per\mol} with respect to a reference composite scheme, which is able to reach sub-kJ accuracy energetics. These errors are much smaller than those issuing from widely employed composite schemes (e.g. CBS-QB3, \cite{CBSQB3}, or G4, \cite{G4}) and well sufficient for obtaining quantitative estimates of reaction rates and branching ratios \citep{earthspace2020,Molecules,lupi:h2s,staa1652,Tonolo2020}. On these grounds, we have performed a full characterization of the doublet PES for the addition-elimination reactions of both CN and CCH radicals to methanimine at the ChS level.

As far as the reaction mechanism is concerned, hydrogen abstraction could be competitive with addition/elimination \citep{Bowman20}, but test computations showed that the former reaction channel is at least one order of magnitude slower than the latter one. As a consequence, only the addition/elimination reaction channel is analyzed in detail in the following. 
The reaction mechanism proposed in the present paper for the formation of N-EMIM and the PGIM isomers is sketched in Figure \ref{fig:pespgimchs} and the relative energies of all the stationary points, with respect to reactants, are collected in Table \ref{tab:cchenergies} together with the corresponding results for the \ce{CH2NH + CN} reaction. 
There are three possible initial adducts, corresponding to the attack of the ethynyl radical to the C or N ends and to the $\pi$-system of the imine double bond. However, the cyclic adduct resulting from the third option (CYCLO-1) is significantly less stable and easily interconverts to one of the corresponding open-chain minima (1Z or 1N). For both the CN and CCH radicals, the intermediate obtained upon attack to the N moiety is slightly more stable, but the reaction channels originating from it are ruled by transition states significantly less stable (albeit always submerged) than those ruling the corresponding channels issuing from 1Z or 1E. Noted is that the PES for the \ce{CH2NH + CN} reaction is, in any detail, analogous to that of the \ce{C2H} radical.

\begin{table}[!htbp]
\centering
\caption{ChS relative electronic energies ($\Delta E_{el}$) and corresponding standard enthalpies at 0 K ($\Delta H^{\circ}_0$) for the stationary points of the \ce{CH2NH + X} reaction. Values in \SI{}{\kilo\joule\per\mol}.}
\label{tab:cchenergies}
\resizebox{0.65\textwidth}{!}{%
\begin{tabular}{@{}ccccc@{}}
\toprule
                     & \multicolumn{2}{c}{X = C$_2$H} & \multicolumn{2}{c}{X = CN} \\
\cline{2-3}\cline{4-5}
                     & $\Delta E_{el}$   & $\Delta H^{\circ}_0$                 & $\Delta E_{el}$    & $\Delta H^{\circ}_0$ \\ \midrule
\ce{CH2NH + X}       & 0.00    & \multicolumn{1}{c|}{0.00}    & 0.00    & 0.00      \\
1Z                   & -229.36 & \multicolumn{1}{c|}{-218.72} & -203.59 & -198.67   \\
TS-1E1Z              & -217.65 & \multicolumn{1}{c|}{-205.90} & -192.39       & -183.24         \\
1E                   & -225.75 & \multicolumn{1}{c|}{-213.58} & -201.48       & -191.95         \\
TS-1Z2               & -96.36  & \multicolumn{1}{c|}{-93.12}  & -63.55  & -62.81    \\
TS-1E2               & -88.80  & \multicolumn{1}{c|}{-86.01}  & -57.47       & -57.18         \\
2                    & -314.60 & \multicolumn{1}{c|}{-299.69} & -284.59 & -271.29   \\
TS-2PZ               & -88.09  & \multicolumn{1}{c|}{-94.97}  & -48.37  & -58.27    \\
TS-2PE               & -84.99  & \multicolumn{1}{c|}{-92.15}  & -46.22  & -56.40    \\
TS-1ZPZ              & -77.81  & \multicolumn{1}{c|}{-83.90}  & -40.62  & -49.89    \\
TS-1EPE              & -73.86  & \multicolumn{1}{c|}{-80.46}  & -37.97  & -47.70    \\
\ce{Z{-}IM + H} (PZ) & -99.21  & \multicolumn{1}{c|}{-110.00} & -60.89  & -74.73    \\
\ce{E{-}IM + H} (PE) & -95.40  & \multicolumn{1}{c|}{-106.61} & -58.55  & -72.78    \\
CYCLO-1              & -167.50 & \multicolumn{1}{c|}{-151.41} & -108.93       & -96.69         \\
TS-CY-C              & -144.35 & \multicolumn{1}{c|}{-133.03} & -97.62       & -88.32         \\
TS-CY-N              & -122.80 & \multicolumn{1}{c|}{-114.02} & -88.77       & -80.94         \\
1N                   & -233.04 & \multicolumn{1}{c|}{-223.67} & -208.44 & -201.32   \\
TS-1N2N              & -53.16  & \multicolumn{1}{c|}{-53.04}  & -25.66  & -27.15    \\
2N                   & -272.23 & \multicolumn{1}{c|}{-260.55} & -223.19 & -211.85   \\
TS-2NPN              & -52.98  & \multicolumn{1}{c|}{-62.22}  & -22.76  & -33.56    \\
TS-1NPN              & -35.54  & \multicolumn{1}{c|}{-44.05}  & -3.85   & -14.67    \\
\ce{N{-}IM + H} (PN) & -59.22  & \multicolumn{1}{c|}{-72.69}  & -30.64  & -45.80    \\ \bottomrule
\end{tabular}%
}
\end{table}

Starting from the very stable 1Z (or 1E) pre-reactive complex, one might observe a loss of the hydrogen radical, leading directly to the Z (or E) isomer of PGIM. This step has an exit barrier of about $\sim$135 (or $\sim$133) kJ mol$^{-1}$. On the other hand, considering the presence of the stabilizing \ce{C2H} moiety on the carbon atom, hydrogen migration might be observed in order to localize the unpaired electron on this atom. This migration occurs through the submerged transition state TS-1Z2 (TS-1E2 for the E-PGIM), which lies 125.6 kJ mol$^{-1}$ above 1Z (127.6 kJ mol$^{-1}$ above 1E for the E-route), thus forming the most stable intermediate of the whole PES, namely 2, which is nearly 300 kJ mol$^{-1}$ below the reactants. Next, loss of hydrogen leads again to the Z (or E) form of PGIM through the submerged transition state TS-2PZ (TS-2PE), lying about 95 kJ mol$^{-1}$ (92 kJ mol$^{-1}$ for the E species) below the reactants (exit barrier of about 205 and 208 kJ mol$^{-1}$, respectively).
The comparison with the analogous reaction paths for the gas-phase production of C-cyanomethanimine \citep{Puzzarini2020} shows that the formation of PGIM is characterized by greater exothermicity (-108 vs. -60 kJ mol$^{-1}$ for the average of Z and E isomers) and lower exit barriers (126 vs. 140 kJ mol$^{-1}$ for the average of TS-1Z2 and TS-1E2 and 206 vs. 238 kJ mol$^{-1}$ for the average of TS-2PZ and TS-2PE). Furthermore, the stability of the pre-reactive complex 1Z or 1E (ruling the barrier-less entrance channel) and that of the intermediate 2 (involved in the two-step mechanism) are greater in the case of the addition of \ce{C2H} than for CN (-218.7 vs. -203.6 kJ mol$^{-1}$ for the average of 1Z,1E and -299.7 vs. -284.6 kJ mol$^{-1}$ for 2). 

Moving to the attack to the N-end of methanimine, from the inspection of Figure~\ref{fig:pespgimchs}, it is evident that the two possible paths originating from the 1N pre-reactive complex are similar to those described above for the C-end attack, as already noted for the \ce{CH2NH + CN} reaction \citep{Vazart2015}. With the only exception of 1N, which lies lower in energy than 1Z and 1E, all intermediates and transition states of these paths are less stable with respect to the C-end counterparts. The product itself, i.e. N-EMIM + H (PN), lies at higher energy: -72.7 kJ mol$^{-1}$, to be compared with -106.6 kJ mol$^{-1}$ for E-PGIM + H (PE) and -110.0 kJ mol$^{-1}$ for Z-PGIM + H (PZ).

Studies for reactions of radical species with molecules containing a double bond have shown that the reactivity depends on the type of system. For the C=C bond, addition/elimination is barrierless and strongly favored over hydrogen elimination (e.g. \cite{jp301015b}), whereas for C=O bonds, only H elimination is barrierless, whereas both the C- and O- addition/eliminations involve small barriers (e.g. \cite{1.1903945}). Preliminary computations for the addition of other radicals (e.g. CP, OH, and \ce{CH3}) to methanimine show that the mechanism described in the previous paragraphs for the reaction with \ce{C2H} or CN represents a quite general route to the formation of complex imines, although in a few cases (e.g. \ce{CH3}) some transition states are not submerged with respect to reactants.

\subsection{Rate constants}

\begin{figure*}[!htbp]
\gridline{\fig{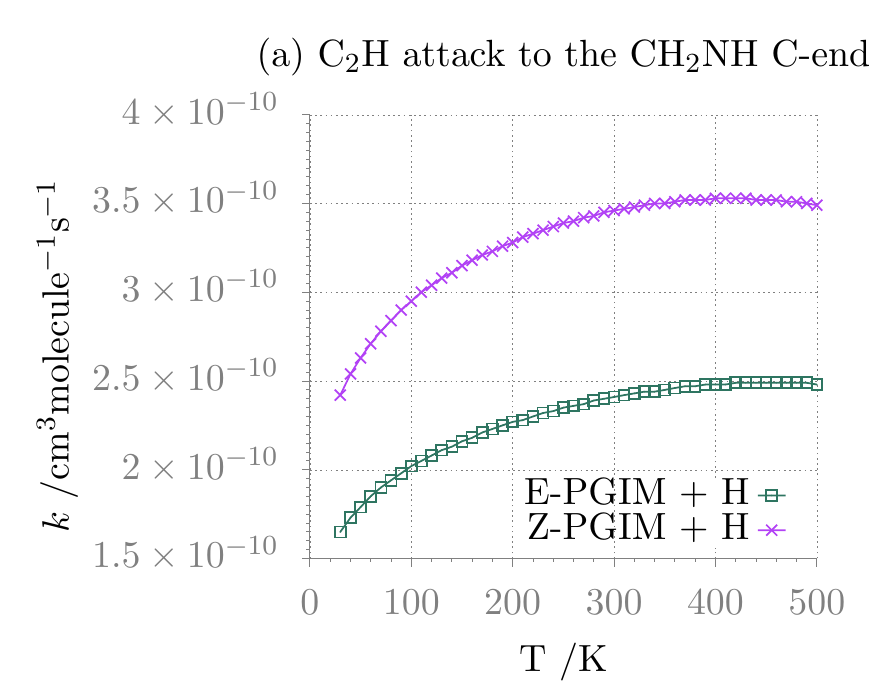}{0.5\textwidth}{}
          \fig{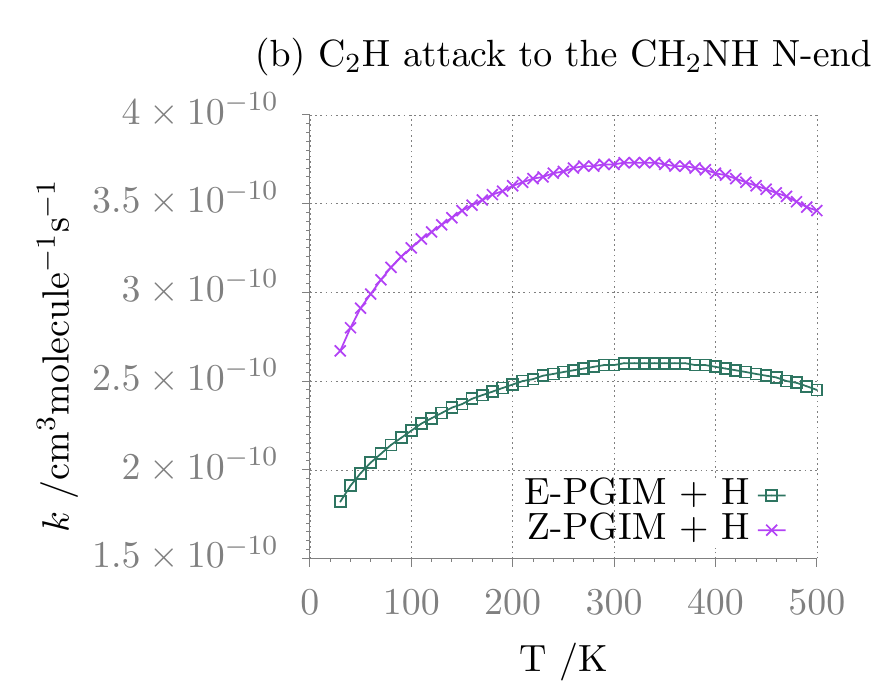}{0.5\textwidth}{}
          }
          \gridline{\fig{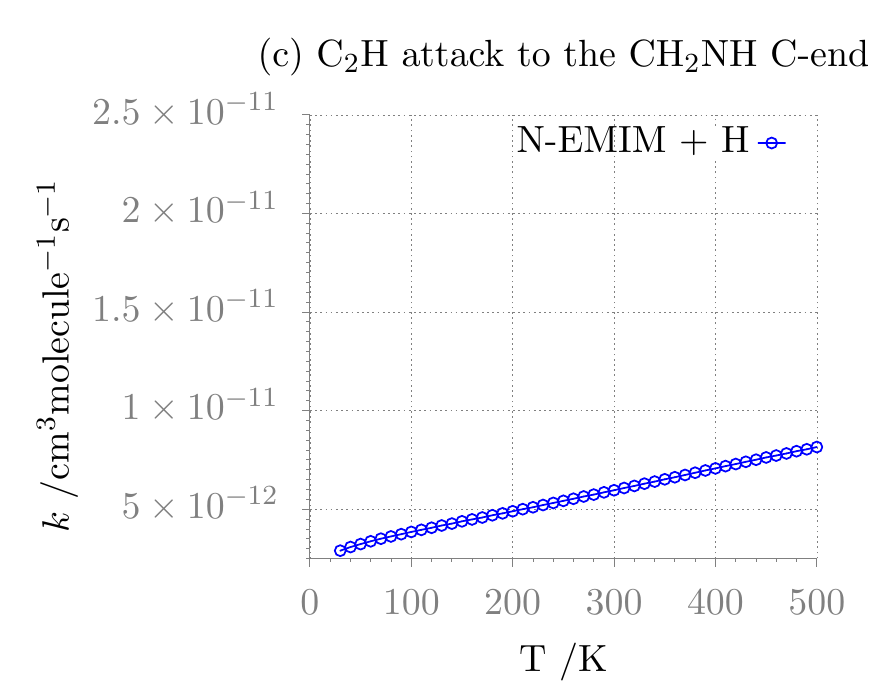}{0.5\textwidth}{}
          \fig{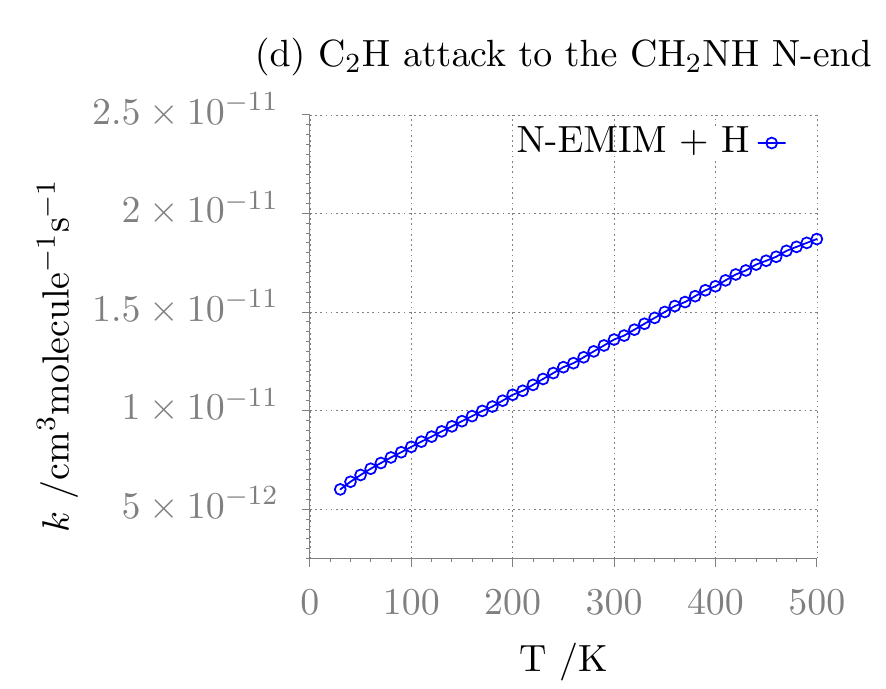}{0.5\textwidth}{}
          }
\caption{Temperature dependence plots of the \ce{CH2NH + C2H} reaction rate constants.
\label{fig:ratepgimchs}}
\end{figure*}

\begin{figure*}[!htbp]
\gridline{\fig{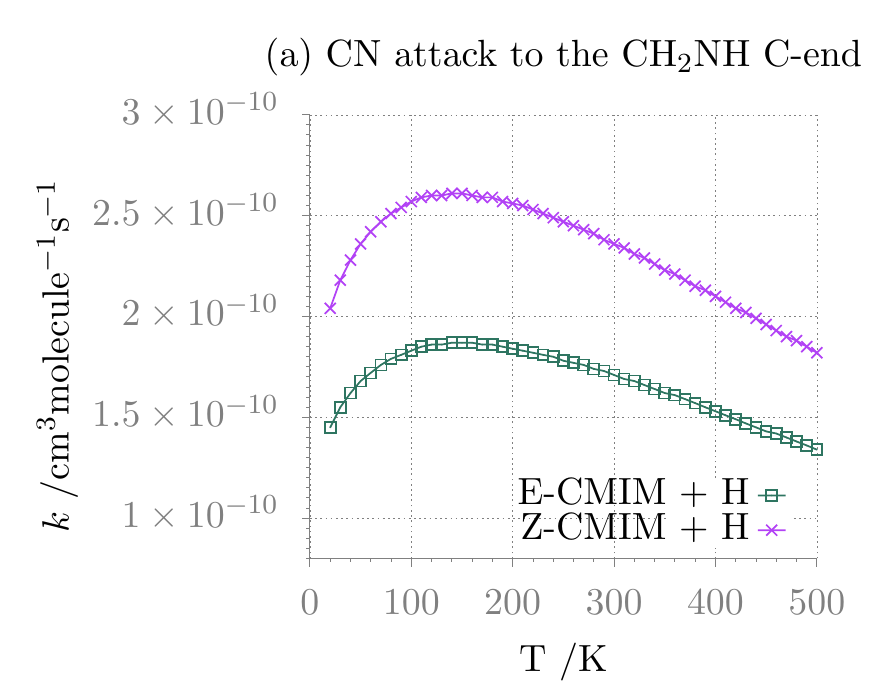}{0.5\textwidth} {}
          \fig{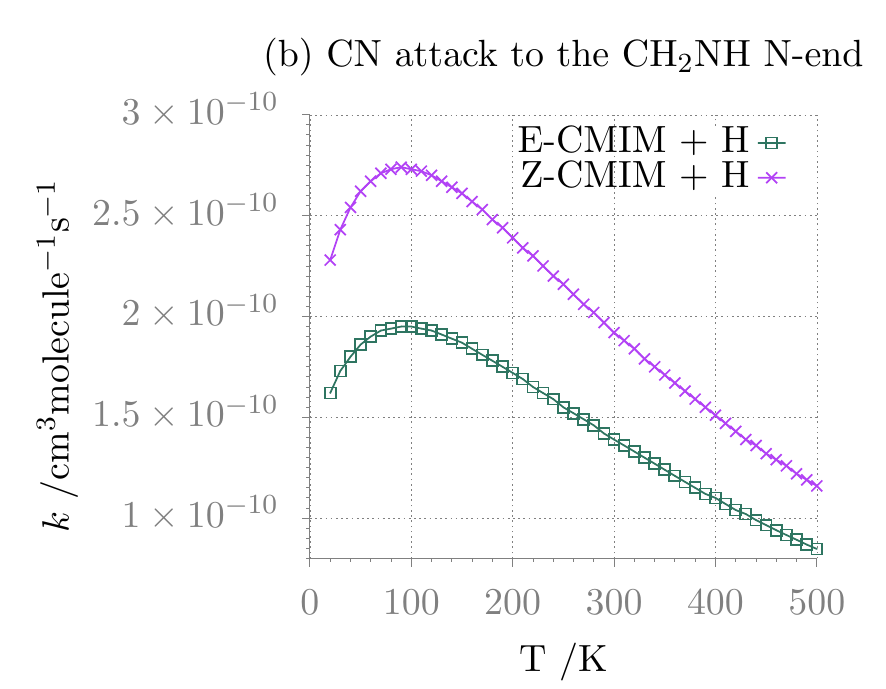}{0.5\textwidth} {}
          }
          \gridline{\fig{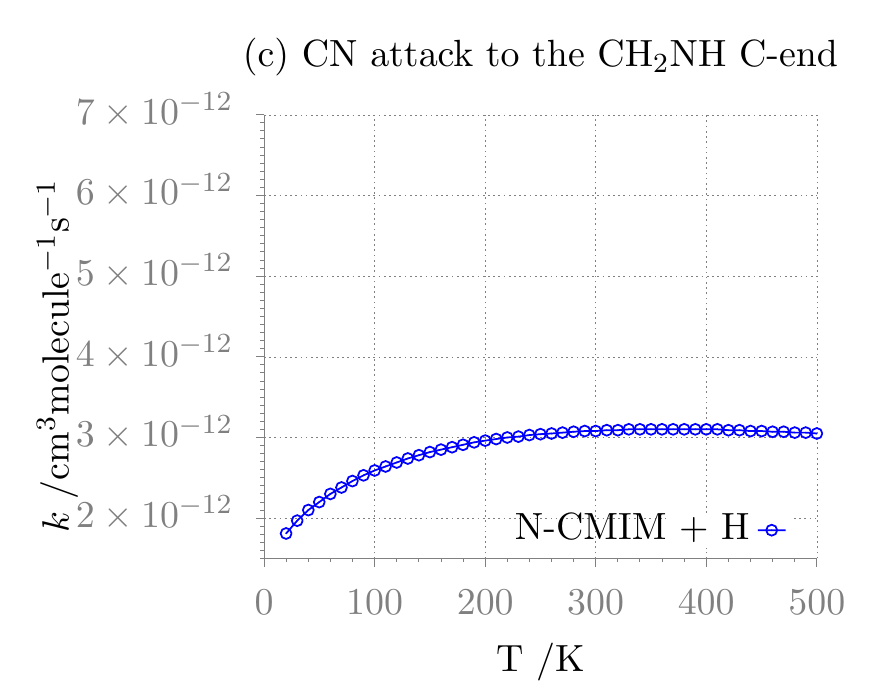}{0.5\textwidth}{}
          \fig{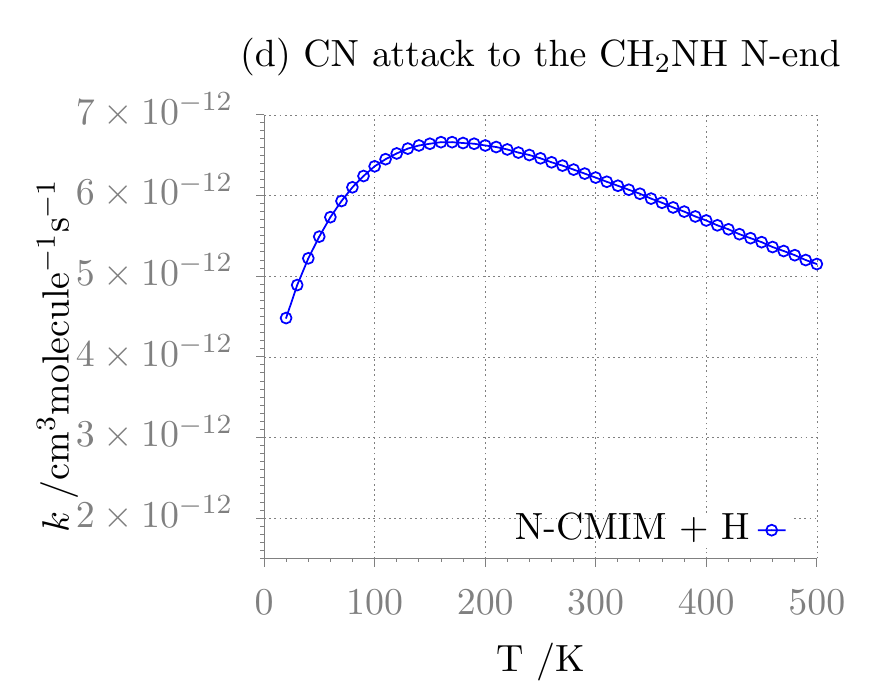}{0.5\textwidth}{}
          }
\caption{Temperature dependence plots of the \ce{CH2NH + CN} reaction rate constants.
\label{fig:ratecmimchs}}
\end{figure*}

To definitely confirm the effectiveness of the mechanism proposed, kinetic computations are required.
The product specific rate constants as a function of temperature are shown in Figure \ref{fig:ratepgimchs} for the reaction of methanimine with \ce{C2H} and in Figure \ref{fig:ratecmimchs} for the reaction with CN, whereas the parameters of the Arrhenius-Kooij fits are given in Table \ref{tab:fitparameters}. These have been obtained by fitting the global rate constants computed in the 20-500 K range. In more detail, for each figure, four panels are provided: those on the left refer to the C-end attack (panels (a) and (c)), while those on the right to the N-end attack (panels (b) and (d)). In both figures, the upper panels show the temperature profiles of rate constants for the formation of the ``C-isomers'' (namely, Z-/E-PGIM and Z-/E-C-cyanomethanimine, CMIM), while the lower panels refer to the formation of the ``N-isomers'' (namely, N-EMIM and N-cyanomethanimine, N-CMIM).

Focusing on the C-end reaction paths, the prevalence of the Z-product is related to the slightly lower energy of the corresponding transition states compared to those leading to the E isomer. Back-dissociation into reactants is negligible in the whole temperature range considered, whereas the overall rate constant for the PGIM formation raises by increasing the temperature, also showing progressive deviations from the Arrhenius behavior. The overall rate constant, which is of the order of 2-3 $\times$ \num{d-10} cm$^3$ molecule$^{-1}$ s$^{-1}$, is mainly ruled by the one-step mechanism leading to products from the 1Z/1E pre-reactive complex through the TS-1ZPZ/TS-1EPE transition state. However, this is always true for Z-PGIM, while for the E isomer the two-step mechanism seems to be the rate determining one above $\sim$350 K. The derived branching ratio is of the order of 1.5, smaller than the observational result ($\ge$1.9), as already noted in the case of C-cyanomethanimine. However, as in the latter case, the isomer abundance ratio obtained from astronomical observations is affected by a large uncertainty (the fractional abundance of E being indirectly derived), but it is, instead, close to the value obtained from a thermodynamic estimate based on the relative stability of the E and Z isomers. The considerations above on the computed isomers ratio assume a similar destruction rate for both E and Z species. In the case of C-cyanomethanimine, it has been suggested that the strong difference between the dipole moments of the E and Z forms (4.2 D and 1.4 D, respectively, from \cite{Vazart2015}) leads to significantly different destruction rates (see \cite{Rivilla20}). More specifically, \cite{Rivilla20} proposed a general rule-of-thumb for estimating the abundances of isomers based on their dipole moments, which has been denoted as ``relative dipole principle''. According to this, for propargylimine, whose isomers have very similar dipole moments ($\sim$2 D, see \cite{Bizzocchi2020}), the assumption that they have similar destruction rates seems to be reliable. At the same time, different reaction rates with H radicals cannot be invoked since, for both cyanomethanimine and propargylimine, the corresponding reactions are ruled by non-submerged transition states. Further investigation of alternative mechanisms would be surely warranted, but it is out of the scope of the present letter. As already noted for thermochemistry, the general kinetic features for the \ce{C2H} and CN additions to methanimine are very similar, thus giving further support to the plausibility and generality of the proposed mechanism. At 100 K, for PGIM, the overall rate constants for Z and E species (in cm$^3$ molecule$^{-1}$ s$^{-1}$) are \num{3.25d-10} and \num{2.22d-10}, respectively, to be compared to \num{2.73d-10} and \num{1.95d-10} for the two corresponding isomers of C-cyanomethanimine. 

\begin{figure*}[!htbp]
\gridline{\fig{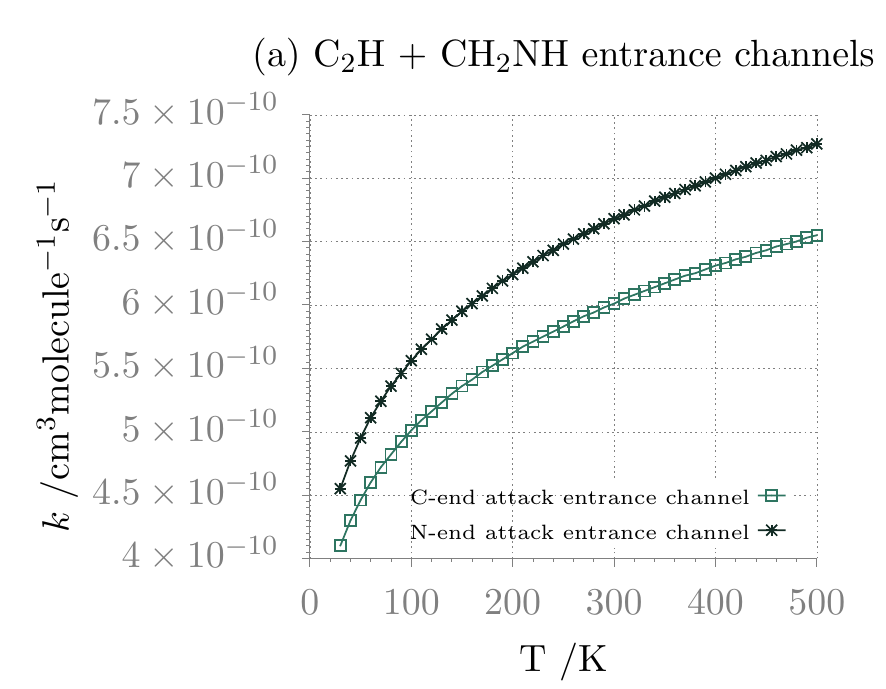}{0.35\textwidth}{}
         \fig{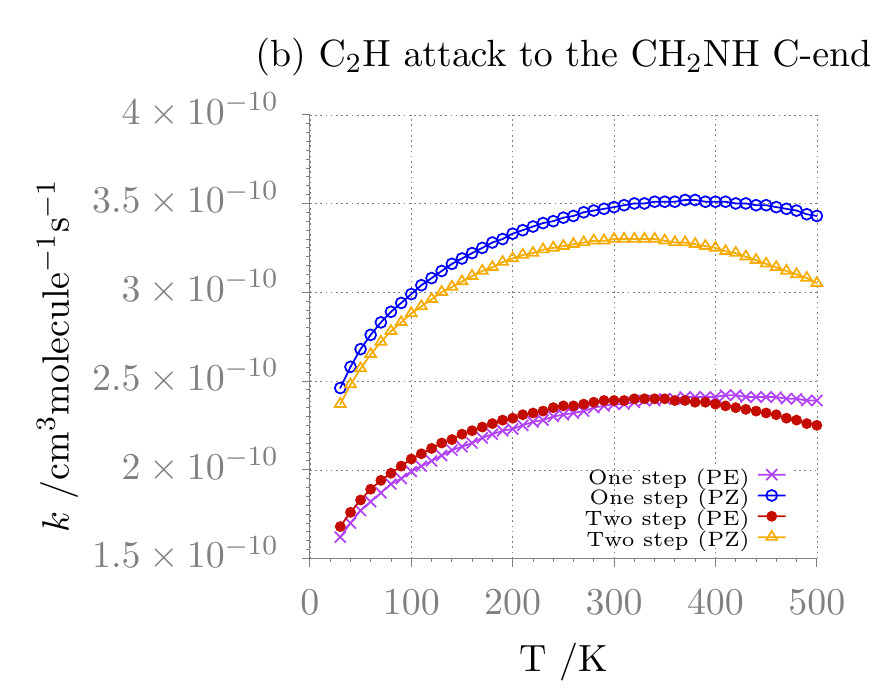}{0.35\textwidth}{}
         \fig{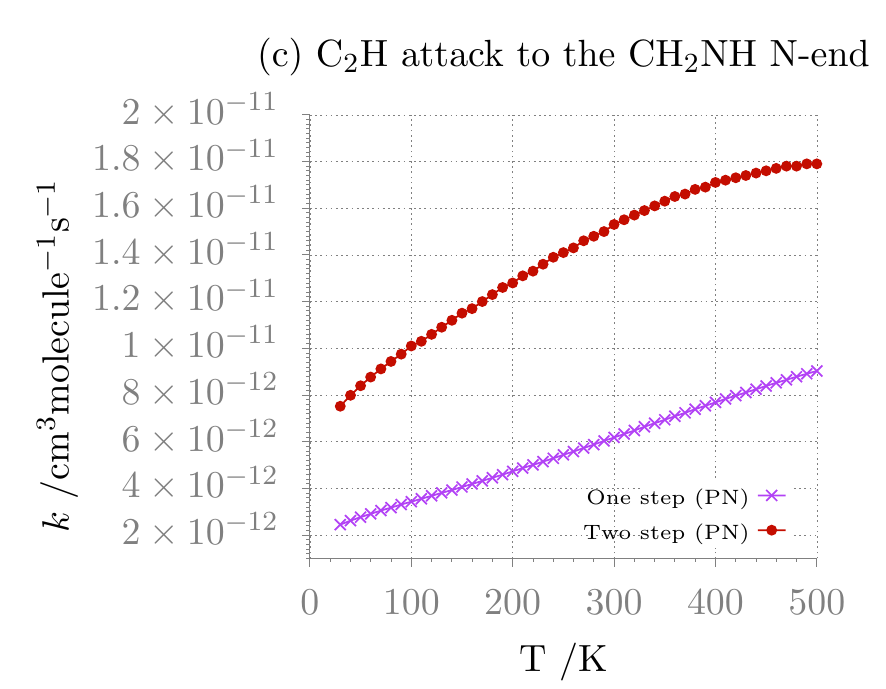}{0.35\textwidth}{}
          }
\caption{Temperature dependence of the rate constants for the elementary steps of the overall \ce{CH2NH + C2H} reaction, namely barrier-less entrance (panel (a)), and one- or two-step paths leading to Z-/E-PGIM (panel (b)) and N-EMIM (panel (c)).
\label{fig:ratecontributipgim}}
\end{figure*}

As far as the formation of the N-species is concerned, it is interesting to note that this process would be a little bit favored over formation of the C-species if the attacks to the two ends of the imino group would be independent (as actually is in the case of the CN addition to the \ce{CH3} or \ce{NH2} moiety of methylamine, see \cite{staa1652}), with a rate constant of 4-\num{5d-10} cm$^3$ molecule$^{-1}$ s$^{-1}$). However, Figure \ref{fig:pespgimchs} shows that the two channels are connected by a low-lying cyclic intermediate. Under these circumstances (also valid for the attack of the CN radical), the formation of C-products becomes faster by two orders of magnitude with respect to formation of the N-product, with the rate constant of the latter process slightly increasing with the temperature.
To provide a graphical explanation of the behavior of the global constant with temperature, the contributions of some specific reaction channels are shown in Figure \ref{fig:ratecontributipgim}. These are the two barrier-less (C- and N-end) entrance channels, the one- and two-step processes leading to Z-/E-PGIM for the C-end attack and the corresponding channel leading to N-EMIM for the attack to the N end of methanimine. 
It is noted that, even if the entrance channel flux for the N-end attack is faster than the C-end attack one, the subsequent high barriers of the N-EMIM formation path slow down the flux, thus resulting in the preferential formation of the E,Z-PGIM, which presents lower lying barriers. In this picture, an important role is played by the TS-CY-N transition state linking 1N to the cyclic pre-reactive complex, CYCLO-1. In fact, this interconversion is the elementary step characterized by the lowest barrier for the N-end side of the overall \ce{CH2NH + C2H} reaction. Similar arguments also apply to the reaction involving CN.

\begin{table}[!htbp]
\centering
\caption{The Arrhenius-Kooij parameters for the \ce{CH2NH + X} reaction.}
\label{tab:fitparameters}
\resizebox{0.9\textwidth}{!}{%
\begin{tabular}{@{}lccc|ccc@{}}
\toprule
\multicolumn{1}{c}{}                 & \multicolumn{3}{c}{C-end attack}                 & \multicolumn{3}{c}{N-end attack}                 \\ \cmidrule(l){2-7} 
X = \ce{C2H}                              & E              & Z              & N              & E              & Z              & N              \\ \cmidrule(l){2-7} 
$A$ /cm$^3$ molecule$^{-1}$ s$^{-1}$ & \num{2.43d-10} & \num{3.51d-10} & \num{5.54d-12} & \num{2.66d-10} & \num{3.83d-10} & \num{1.26d-11} \\
$n$                                  & \num{7.58d-2}  & \num{3.86d-2}  & \num{6.33d-1}  & \num{-6.10d-2} & \num{-9.22d-2} & \num{6.59d-1}  \\
$E$ /\SI{}{\kilo\joule\per\mol}      & \num{6.74d-2}  & \num{8.72d-2}  & \num{-2.32d-1} & \num{1.62d-1}  & \num{1.77d-1}  & \num{-2.15d-1} \\
rms\textsuperscript{\emph{a}}        & \num{4.37d-12} & \num{7.12d-12} & \num{1.15d-13} & \num{1.02d-11} & \num{1.54d-11} & \num{1.97d-13} \\ \midrule
X = \ce{CN}                               & E              & Z              & N              & E              & Z              & N              \\ \cmidrule(l){2-7} 
$A$ /cm$^3$ molecule$^{-1}$ s$^{-1}$ & \num{1.75d-10} & \num{2.42d-10} & \num{3.12d-12} & \num{1.46d-10} & \num{2.03d-10} & \num{6.51d-12} \\
$n$                                  & \num{-3.20d-1} & \num{-3.40d-1} & \num{2.68d-2}  & \num{-6.56d-1} & \num{-6.68d-1} & \num{-2.72d-1} \\
$E$ /\SI{}{\kilo\joule\per\mol}      & \num{2.17d-1}  & \num{2.24d-1}  & \num{1.03d-1}  & \num{3.37d-1}  & \num{3.39d-1}  & \num{2.33d-1}  \\
rms\textsuperscript{\emph{a}}        & \num{1.09d-11} & \num{1.55d-11} & \num{8.39d-14} & \num{1.47d-11} & \num{2.07d-11} & \num{3.29d-13} \\ \bottomrule
\end{tabular}%
}

\textsuperscript{\emph{a}} rms stands for root-mean-square deviation of the fit.
\end{table}

It is noteworthy that the behavior discussed above for both types of radical attack to methanimine is specific of the low pressure limit (see computational details). In fact, moving to a pressure of 1 atm (of limited astrophysical interest, but of potential relevance in  planetary atmospheres), the N-EMIM formation remains unfavorable with respect to E,Z-PGIM and, in general, all formation rate constants become slower. This trend is due to the stabilization of the entrance channel wells (namely 1N and 1Z) by collisions (that occurs at pressure values as high as 1 atm), thus leading to an increase of the effective reaction barriers with the consequent decrease of the overall rate constant, which shows a monotonic increase with temperature.

A curved Arrhenius plot is obtained when the activation energy depends on the temperature and this behavior is captured by the Arrhenius-Kooij formula (see Equation \ref{eq:kooij}) when this dependence is linear. The root mean square deviations reported in Table \ref{tab:fitparameters} demonstrate that the data for the \ce{C3H3N} imine isomers are indeed well fitted by the Arrhenius-Kooij expression. Within this model, $E$ represents the activation energy at 0 K and the activation energy at a generic temperature $T$ is given by $E+n\left(\frac{RT}{300}\right)$. In the present case, the activation energy is always positive, with the exception of N-EMIM, as a result of both the capture rate constant and the subsequent energy barriers for the unimolecular steps. The $n$ parameter (the first derivative of the activation energy with respect to temperature) is always positive for the C-end attack, while it is negative for the PGIM isomers when the N-end attack takes place. Finally, the values of the pre-exponential factor $A$ are typical for this kind of reactions and rule the branching ratio between the Z and E PGIM isomers. Indeed, the ratio of the $A$ factors is 1.44 and the branching ratio ranges between 1.43 and 1.47 in the whole temperature range (20-500 K).

\section{Concluding remarks}  
In this letter, we have proposed a gas-phase formation route for the recently detected Z-PGIM molecule.  
In analogy to the addition of the CN radical to methanimine leading to cyanomethanimine, addition of the isoelectronic ethynyl radical easily leads to PGIM through a similar reaction mechanism, which involves the formation of a stable pre-reactive complex and its successive evolution by means of submerged transition states. 
Since the level of the QC and kinetic computations carried out gives strong supports to the quantitative accuracy of our results, search for PGIM isomers in the other regions of the ISM where methanimine and the ethynyl radical have been both detected could be attempted to further validate the proposed reaction mechanism.

In a more general perspective, the results of our state-of-the-art computations provide convincing evidences about the feasibility of a general addition/elimination mechanism for the formation of complex imines. This starts from methanimine as a precursor and involves reactive radicals abundantly present in the interstellar space.

\acknowledgments
This work has been supported by MIUR 
(Grant Number 2017A4XRCA) and by the University of Bologna (RFO funds). The SMART@SNS Laboratory (http://smart.sns.it) is acknowledged for providing high-performance computing facilities. Support by the Italian Space Agency (ASI; `Life in Space' project, N. 2019-3-U.0) is also acknowledged.

%




\clearpage

\bibliography{Astrochem}{}

\begin{thebibliography}{}
\expandafter\ifx\csname natexlab\endcsname\relax\def\natexlab#1{#1}\fi
\providecommand{\url}[1]{\href{#1}{#1}}
\providecommand{\dodoi}[1]{doi:~\href{http://doi.org/#1}{\nolinkurl{#1}}}
\providecommand{\doeprint}[1]{\href{http://ascl.net/#1}{\nolinkurl{http://ascl.net/#1}}}
\providecommand{\doarXiv}[1]{\href{https://arxiv.org/abs/#1}{\nolinkurl{https://arxiv.org/abs/#1}}}

\bibitem[{Baiano {et~al.}(2020)Baiano, Lupi, Tasinato, Puzzarini, \&
  Barone}]{Molecules}
Baiano, C., Lupi, J., Tasinato, N., Puzzarini, C., \& Barone, V. 2020,
  Molecules, 25, 2873

\bibitem[{Bizzocchi {et~al.}(2020)Bizzocchi, {Prudenzano, D.}, {Rivilla, V.
  M.}, {Pietropolli-Charmet, A.}, {Giuliano, B. M.}, {Caselli, P.},
  {Mart\'{\i}n-Pintado, J.}, {Jim\'enez-Serra, I.}, {Mart\'{\i}n, S.},
  {Requena-Torres, M. A.}, {Rico-Villas, F.}, {Zeng, S.}, \& {Guillemin,
  J.-C.}}]{Bizzocchi2020}
Bizzocchi, L., {Prudenzano, D.}, {Rivilla, V. M.}, {et~al.} 2020, Astron.
  Astrophys., 640, A98

\bibitem[{Bloino {et~al.}(2012)Bloino, Biczysko, \& Barone}]{Bloino2012}
Bloino, J., Biczysko, M., \& Barone, V. 2012, J. Chem. Theory Comput., 8, 1015

\bibitem[{Bouwman {et~al.}(2012)Bouwman, Goulay, Leone, \& Wilson}]{jp301015b}
Bouwman, J., Goulay, F., Leone, S.~R., \& Wilson, K.~R. 2012, J. Phys. Chem. A,
  116, 3907

\bibitem[{Bowman {et~al.}(2020)Bowman, Burke, Turney, \&
  Schaefer~III}]{Bowman20}
Bowman, M.~C., Burke, A.~D., Turney, J.~M., \& Schaefer~III, H.~F. 2020, Mol.
  Phys., 118, in press. DOI: 10.1080/00268976.2020.1769214

\bibitem[{Chesnavich(1986)}]{chesnavich1986multiple}
Chesnavich, W., J. 1986, J. Chem. Phys., 84, 2615

\bibitem[{Codella {et~al.}(2017)Codella, {Ceccarelli, C.}, {Caselli, P.},
  {Balucani, N.}, {Barone, V.}, {Fontani, F.}, {Lefloch, B.}, {Podio, L.},
  {Viti, S.}, {Feng, S.}, {Bachiller, R.}, {Bianchi, E.}, {Dulieu, F.},
  {Jim\'enez-Serra, I.}, {Holdship, J.}, {Neri, R.}, {Pineda, J. E.}, {Pon,
  A.}, {Sims, I.}, {Spezzano, S.}, {Vasyunin, A. I.}, {Alves, F.}, {Bizzocchi,
  L.}, {Bottinelli, S.}, {Caux, E.}, {Chac\'on-Tanarro, A.}, {Choudhury, R.},
  {Coutens, A.}, {Favre, C.}, {Hily-Blant, P.}, {Kahane, C.}, {Jaber Al-Edhari,
  A.}, {Laas, J.}, {L\'opez-Sepulcre, A.}, {Ospina, J.}, {Oya, Y.}, {Punanova,
  A.}, {Puzzarini, C.}, {Quenard, D.}, {Rimola, A.}, {Sakai, N.}, {Skouteris,
  D.}, {Taquet, V.}, {Testi, L.}, {Theul\'e, P.}, {Ugliengo, P.}, {Vastel, C.},
  {Vazart, F.}, {Wiesenfeld, L.}, \& {Yamamoto, S.}}]{formamide-solis}
Codella, C., {Ceccarelli, C.}, {Caselli, P.}, {et~al.} 2017, Astronomy
  Astrophys., 605, L3

\bibitem[{Curtiss {et~al.}(2007)Curtiss, Redfern, \& Raghavachari}]{G4}
Curtiss, L., A., Redfern, P., C., \& Raghavachari, K. 2007, J. Chem. Phys.,
  126, 084108

\bibitem[{Dickens {et~al.}(1997)Dickens, Irvine, DeVries, \&
  Ohishi}]{Dickens_1997}
Dickens, J.~E., Irvine, W.~M., DeVries, C.~H., \& Ohishi, M. 1997, Astrophys.
  J., 479, 307

\bibitem[{Dong {et~al.}(2005)Dong, Ding, \& Sun}]{1.1903945}
Dong, H., Ding, Y.-h., \& Sun, C.-c. 2005, J. Chem. Phys., 122, 204321

\bibitem[{Dunning~Jr.(1989)}]{Dunning-JCP1989_cc-pVxZ}
Dunning~Jr., T.~H. 1989, J. Chem. Phys., 90, 1007

\bibitem[{Eckart(1930)}]{eckart1930penetration}
Eckart, C. 1930, Phys. Rev., 35, 1303

\bibitem[{Fernández-Ramos {et~al.}(2006)Fernández-Ramos, Miller,
  Klippenstein, \& Truhlar}]{doi:10.1021/cr050205w}
Fernández-Ramos, A., Miller, J.~A., Klippenstein, S.~J., \& Truhlar, D.~G.
  2006, Chem. Rev., 106, 4518

\bibitem[{Frisch {et~al.}(2016)Frisch, Trucks, Schlegel, Scuseria, Robb,
  Cheeseman, Scalmani, Barone, Petersson, Nakatsuji, Li, Caricato, Marenich,
  Bloino, Janesko, Gomperts, Mennucci, Hratchian, Ortiz, Izmaylov, Sonnenberg,
  Williams-Young, Ding, Lipparini, Egidi, Goings, Peng, Petrone, Henderson,
  Ranasinghe, Zakrzewski, Gao, Rega, Zheng, Liang, Hada, Ehara, Toyota, Fukuda,
  Hasegawa, Ishida, Nakajima, Honda, Kitao, Nakai, Vreven, Throssell,
  Montgomery, Peralta, Ogliaro, Bearpark, Heyd, Brothers, Kudin, Staroverov,
  Keith, Kobayashi, Normand, Raghavachari, Rendell, Burant, Iyengar, Tomasi,
  Cossi, Millam, Klene, Adamo, Cammi, Ochterski, Martin, Morokuma, Farkas,
  Foresman, \& Fox}]{g16}
Frisch, M.~J., Trucks, G., W., Schlegel, H., B., {et~al.} 2016, Gaussian 16
  {R}evision {C}.01

\bibitem[{Georgievskii {et~al.}(2013)Georgievskii, Miller, Burke, \&
  Klippenstein}]{georgievskii2013reformulation}
Georgievskii, Y., Miller, J., A., Burke, M., P., \& Klippenstein, S., J. 2013,
  J. Phys. Chem. A, 117, 12146

\bibitem[{Godfrey {et~al.}(1973)Godfrey, Brown, Robinson, \&
  Sinclair}]{godfrey73}
Godfrey, P.~D., Brown, R.~D., Robinson, B.~J., \& Sinclair, M.~W. 1973,
  Astrophys. Lett., 13, 119

\bibitem[{Grimme(2006)}]{doi:grimme2006}
Grimme, S. 2006, J. Chem. Phys., 124, 034108

\bibitem[{Grimme {et~al.}(2010)Grimme, Antony, Ehrlich, \& Krieg}]{D3}
Grimme, S., Antony, J., Ehrlich, S., \& Krieg, H. 2010, J. Chem. Phys., 132,
  154104

\bibitem[{Grimme {et~al.}(2011)Grimme, Ehrlich, \& Goerigk}]{D3BJ}
Grimme, S., Ehrlich, S., \& Goerigk, L. 2011, J. Comp. Chem., 32, 1456

\bibitem[{Helgaker {et~al.}(1997)Helgaker, Klopper, Koch, \&
  Noga}]{Helgaker1997}
Helgaker, T., Klopper, W., Koch, H., \& Noga, J. 1997, J. Chem. Phys., 106,
  9639

\bibitem[{Herbst \& van Dishoeck(2009)}]{COMs}
Herbst, E., \& van Dishoeck, E., F. 2009, Ann. Rev. Astron. Astrophys., 47, 427

\bibitem[{Kawaguchi {et~al.}(1992)Kawaguchi, Takano, Ohishi, Ishikawa,
  Miyazawa, Kaifu, Yamashita, Yamamoto, Saito, Ohshima,
  {et~al.}}]{kawaguchi1992detection}
Kawaguchi, K., Takano, S., Ohishi, M., {et~al.} 1992, Astrophys. J., 396, L49

\bibitem[{Kooij(1893)}]{kooij1893zersetzung}
Kooij, D. 1893, Zeitschr. Phys. Chem., 12, 155

\bibitem[{Krim {et~al.}(2019)Krim, Guillemin, \& Woon}]{Krim2019}
Krim, L., Guillemin, J., C., \& Woon, D., E. 2019, MNRAS, 485, 5210

\bibitem[{Laidler(1996)}]{laidler1996glossary}
Laidler, K.~J. 1996, Pure Appl. Chem., 68, 149

\bibitem[{Loomis {et~al.}(2013)Loomis, Zaleski, Steber, Neill, Muckle, Harris,
  Hollis, Jewell, Lattanzi, Lovas, {et~al.}}]{loomis2013detection}
Loomis, R.~A., Zaleski, D.~P., Steber, A.~L., {et~al.} 2013, Astrophys. J.,
  765, L9

\bibitem[{Lovas {et~al.}(2006)Lovas, Hollis, Remijan, \& Jewell}]{Lovas_2006}
Lovas, F.~J., Hollis, J.~M., Remijan, A.~J., \& Jewell, P.~R. 2006, Astrophys.
  J., 645, L137

\bibitem[{Lupi {et~al.}(2020)Lupi, Puzzarini, Cavallotti, \& Barone}]{lupi:h2s}
Lupi, J., Puzzarini, C., Cavallotti, C., \& Barone, V. 2020, J. Chem. Theory
  Comput., 16, 5090

\bibitem[{McGuire(2018)}]{McGuire_2018}
McGuire, B., A. 2018, Astrophys. J. Suppl. Ser., 239, 17

\bibitem[{M{\o}ller \& Plesset(1934)}]{mp2}
M{\o}ller, C., \& Plesset, M.~S. 1934, Phys. Rev., 46, 618

\bibitem[{Montgomery {et~al.}(2000)Montgomery, Frisch, Ochterski, \&
  Petersson}]{CBSQB3}
Montgomery, J.~A., Frisch, M.~J., Ochterski, J.~W., \& Petersson, G.~A. 2000,
  J. Chem. Phys., 112, 6532

\bibitem[{Papajak {et~al.}(2009)Papajak, Leverentz, Zheng, \&
  Truhlar}]{papajak2009}
Papajak, E., Leverentz, H., R., Zheng, J., \& Truhlar, D.~G. 2009, J. Chem.
  Theory Comput., 5, 1197

\bibitem[{Pechukas \& Light(1965)}]{pechukas1965detailed}
Pechukas, P., \& Light, J., C. 1965, J. Chem. Phys., 42, 3281

\bibitem[{Puzzarini {et~al.}(2014)Puzzarini, , Biczysko, Barone, Largo, Pena,
  Cabezas, \& Alonso}]{cheap2}
Puzzarini, C., , Biczysko, M., {et~al.} 2014, J. Phys. Chem. Lett., 5, 534

\bibitem[{Puzzarini \& Barone(2011)}]{cheap1}
Puzzarini, C., \& Barone, V. 2011, Phys. Chem. Chem. Phys., 13, 7180

\bibitem[{Puzzarini \& Barone(2020)}]{Puzzarini2020}
---. 2020, Phys. Chem. Chem. Phys., 22, 6507

\bibitem[{Puzzarini {et~al.}(2019)Puzzarini, Bloino, Tasinato, \&
  Barone}]{chemRev2019}
Puzzarini, C., Bloino, J., Tasinato, N., \& Barone, V. 2019, Chem. Rev., 119,
  8131

\bibitem[{Puzzarini {et~al.}(2020)Puzzarini, Salta, Tasinato, Lupi, Cavallotti,
  \& Barone}]{staa1652}
Puzzarini, C., Salta, Z., Tasinato, N., {et~al.} 2020, Mon. Not. R. Astron.
  Soc., 496, 4298

\bibitem[{Raghavachari {et~al.}(1989)Raghavachari, Trucks, Pople, \&
  Head-Gordon}]{Pople89}
Raghavachari, K., Trucks, G.~W., Pople, J.~A., \& Head-Gordon, M. 1989, Chem.
  Phys. Lett., 157, 479

\bibitem[{Rivilla {et~al.}(2018)Rivilla, Martín-Pintado, Jiménez-Serra, Zeng,
  Martín, Armijos-Abendaño, Requena-Torres, Aladro, \&
  Riquelme}]{rivilla2019}
Rivilla, V.~M., Martín-Pintado, J., Jiménez-Serra, I., {et~al.} 2018, Mon.
  Not. R. Astron. Soc., 483, L114

\bibitem[{Salta {et~al.}(2020)Salta, Tasinato, Lupi, Boussessi, Balbi,
  Puzzarini, \& Barone}]{earthspace2020}
Salta, Z., Tasinato, N., Lupi, J., {et~al.} 2020, ACS Earth Space Chem., 4, 774

\bibitem[{Shingledecker {et~al.}(2020)Shingledecker, Molpeceres, Rivilla,
  Majumdar, \& K{\"a}stner}]{Rivilla20}
Shingledecker, C.~N., Molpeceres, G., Rivilla, V.~M., Majumdar, L., \&
  K{\"a}stner, J. 2020, Astrophys. J., 897, 158,
  \dodoi{10.3847/1538-4357/ab94b5}

\bibitem[{Theule {et~al.}(2011)Theule, Borget, \& Mispelaer}]{Theule2011}
Theule, P., Borget, F., \& Mispelaer, F. e.~a. 2011, Astron. Astrophys., 534,
  A64

\bibitem[{Tonolo {et~al.}(2020)Tonolo, Lupi, Puzzarini, \& Barone}]{Tonolo2020}
Tonolo, F., Lupi, J., Puzzarini, C., \& Barone, V. 2020, Astrophys. J., 900, 85

\bibitem[{Vazart {et~al.}(2016)Vazart, Calderini, Puzzarini, Skouteris, \&
  Barone}]{Vazart2016}
Vazart, F., Calderini, D., Puzzarini, C., Skouteris, D., \& Barone, V. 2016, J.
  Chem. Theory Comput., 12, 5385

\bibitem[{Vazart {et~al.}(2015)Vazart, Latouche, Skouteris, Balucani, \&
  Barone}]{Vazart2015}
Vazart, F., Latouche, C., Skouteris, D., Balucani, N., \& Barone, V. 2015,
  Astrophys. J., 810, 111

\bibitem[{Woon \& Dunning~Jr.(1995)}]{cvbasis}
Woon, D., E., \& Dunning~Jr., T.~H. 1995, J. Chem. Phys., 103, 4572

\bibitem[{Zaleski {et~al.}(2013)Zaleski, Seifert, Steber, Muckle, Loomis,
  Corby, Martinez~Jr, Crabtree, Jewell, Hollis,
  {et~al.}}]{zaleski2013detection}
Zaleski, D.~P., Seifert, N.~A., Steber, A.~L., {et~al.} 2013, Astrophys. J.,
  765, L10

\bibitem[{{Zeng} {et~al.}(2018){Zeng}, {Jim{\'e}nez-Serra}, {Rivilla},
  {Mart{\'\i}n}, {Mart{\'\i}n-Pintado}, {Requena-Torres},
  {Armijos-Abenda{\~n}o}, {Riquelme}, \& {Aladro}}]{Zeng2018}
{Zeng}, S., {Jim{\'e}nez-Serra}, I., {Rivilla}, V.~M., {et~al.} 2018, Mon. Not.
  R. Astron. Soc., 478, 2962

\end{thebibliography}
\bibliographystyle{aasjournal}



\end{document}